\documentclass[sigconf]{acmart}
\AtBeginDocument{%
  }

\copyrightyear{2026}
\acmYear{2026}
\setcopyright{cc}
\setcctype{by-nc-nd}
\acmConference[RecSys '26]{20th ACM Conference on Recommender Systems}{September 27-October 02, 2026}{Minneapolis, MN, USA}
\acmBooktitle{20th ACM Conference on Recommender Systems (RecSys '26), September 27-October 02, 2026, Minneapolis, MN, USA}
\acmDOI{10.1145/3773078.3841289}
\acmISBN{979-8-4007-2284-4/2026/09}

\acmISBN{978-1-4503-XXXX-X/2018/06}

\usepackage{soul}
\usepackage{mdframed}
\usepackage{subcaption}
\usepackage{stfloats}

\begin{document}

\title{Training seeds and model-selection stability in recommender-system evaluation}

\author{Juan Manuel Rodriguez}
\email{jmro@cs.aau.dk}
\orcid{orcid.org/0000-0002-1130-8065}
\affiliation{%
  \institution{Aalborg University}
  \city{Aalborg}
  \country{Denmark}
}

\author{Oleg Lesota}
\email{oleg.lesota@jku.at}
\orcid{0000-0002-8321-6565}
\affiliation{%
  \institution{Johannes Kepler University Linz}
  \city{Linz}
  \country{Austria}
}

\author{Antonela Tommasel}
\email{antonela.tommasel@jku.at}
\orcid{0000-0001-6091-8305}
\affiliation{%
  \institution{Johannes Kepler University Linz}
  \city{Linz}
  \country{Austria}
}
\affiliation{
  \institution{ISISTAN, CONICET-UNCPBA}
  \city{Tandil}
  \country{Argentina}
}
\renewcommand{\shortauthors}{Rodriguez et al.}

\begin{abstract}
Recommender-system experiments often rely on a single random training seed, assuming that run-to-run stochasticity has limited impact on evaluation conclusions. 
This assumption is risky, as a training seed may influence several algorithm-dependent mechanisms, including parameter initialization, mini-batch ordering, dropout, masking, latent sampling, and training-time negative sampling. 
We examine this assumption by fixing the data partition and varying the training seed across hyperparameter configurations. 
We analyze seed effects at three levels: user-level metric sensitivity, validation-based model selection and recommendation-list agreement.
Results show that seed variation is often detectable
.
Its impact depends on whether configurations are clearly separated, whether validation results transfer to test, and whether similar scores lead to similar top-$k$ lists.
Findings suggest that reporting single-seed results can overstate the stability of recommender system evaluation, 
and that training seeds should be treated as part of the evaluation protocol rather than as incidental implementation noise.
\end{abstract}

\maketitle

\vspace{-0.0cm}\section{Introduction}
\vspace{-0.1cm}Recent reproducibility studies have questioned how confidently progress can be claimed in recommender system research. 
Prior work has shown that results may depend on baseline tuning, implementation details, data splitting, sampled evaluation, model initialization, and training-time sampling choices~\cite{ferrari2019we,d2022analyzing,wegmeth2023effect,krichene2020sampled,prakash2024evaluating}.
However, even under a fixed data partition, hyperparameters, and evaluation protocol, changing the training seed may affect both reported metrics and model-selection conclusions.

This paper focuses on \textit{training-seed variation under an otherwise fixed experimental setup}.
We fix the data partition and vary a single training seed across hyperparameter configurations. 
We interpret the seed as controlling the stochastic components of the full training run, rather than only parameter initialization. Our goal is not to attribute variability to a specific internal mechanism, but to examine whether repeated training-seed changes the conclusions researchers would draw from an experiment.

We examine whether training-seed variation affects user-level evaluation results, whether comparisons observed on validation data are preserved on test data, whether the ranking of hyperparameter configurations remains stable, and how much performance is lost when selecting a configuration based on validation results. 
This framing distinguishes detectable variation from practical instability. A seed may change user-level metric distributions without changing the selected configuration, while near-tied validation results may still affect model selection.

\vspace{-0.25cm}\section{Methodology}
\vspace{-0.1cm}Our methodology isolates training-seed variability under a fixed data partition and evaluates its effect on performance estimates, model selection decisions and recommendation lists\footnote{More details, code and results: \url{https://github.com/knife982000/RecSeed}}.

\vspace{-0.2cm}\paragraph{Datasets.} 
We use three benchmark datasets: Movielens-1M \cite{harper2015movielens}, Steam \cite{SASRec}, and Amazon All Beauty 2023~\cite{amazon}. 
Interactions are ordered temporally and split by user into train, validation and test sets using an $80\%/10\%/10\%$ split. 

\vspace{-0.2cm}\paragraph{Recommendation models.} 
We evaluate four widely used recommender models, covering both matrix-factorization and sequential neural approaches: BPR~\cite{bpr}, NeuMF~\cite{neumf}, BERT4Rec~\cite{BERT4Rec}, and SASRec~\cite{SASRec}. 
BPR uses $3$ embedding-size configuration. NeuMF uses $9$ combinations of MF embedding size and MLP hidden layers. BERT4Rec and SASRec each use $12$ configurations varying hidden layer size, number of layers, and attention heads\footnote{Full hyperparameter spaces are provided in the companion repository.}. 

\vspace{-0.2cm}\paragraph{Training procedure.} 
All models are trained for up to $300$ epochs for each configuration using early stopping with a patience of $10$ epochs.
For each model--configuration pair, we repeat training with 
the fixed seed set $\{1,\ldots,10\}$ (chosen a priori and not based on performance to avoid seed tuning) and retain the checkpoint with the best validation performance. 
Unless otherwise specified, all remaining hyperparameters are kept at their RecBole defaults~\cite{recbole}. 

\vspace{-0.2cm}\paragraph{Evaluation focus.}
We analyze training-seed variation at four complementary levels, using 
nDCG@10 as the primary ranking metric\footnote{Recall@10, MRR@10, Hit@10, and Precision@10 can be found in the repository.}.

\noindent\ul{Metric-level sensitivity.}
For each model and hyperparameter configuration, we compute user-level recommendation scores for each training seed. Since the same users are evaluated across seeds, we treat users as repeated observations and apply the Friedman test to compare the per-user metric distributions obtained under different seeds. 
The analysis captures whether training-seed variation is detectable before considering its effects on model-selection decisions.

\noindent\ul{Model-selection stability.}
We evaluate whether seed variation affects the configuration selected from validation results. 
For each training seed, we identify the best validation configuration and compare it with the best observed under the corresponding evaluation setting. 
We report the performance gap (\%) for validation--validation and validation--test comparisons, 
distinguishing instability within validation from that transferred to the final test evaluation. 

\noindent\ul{Validation--test consistency.}
We evaluate whether pairwise configuration comparisons on validation are preserved on test. For each configuration pair, we estimate how often validation wins transfer to test. This indicates whether validation provides a reliable signal for ranking configurations, beyond their performance gap.

\noindent\ul{Recommendation-list agreement.}
We examine whether different seeds produce similar top-$k$ recommendation lists.
For each configuration, we compare lists generated by pairs of seeds using Jaccard similarity and Average Overlap@$k$: $AO@k = \frac{1}{k}\sum_{i=1}^{k}\frac{|R_1@i \cap R_2@i|}{i}$, where $R_x@i$ are the top-$i$ recommendations by seed $x$.
This captures whether seeds affect the recommendations actually produced.

\vspace{-0.25cm}\section{Experimental results}

\begin{figure*}[t]
    \centering
    \begin{minipage}[t]{0.495\textwidth}
    \vspace{0pt}%
    \centering
    \begin{subfigure}{1\linewidth}
    \includegraphics[width=1\linewidth]{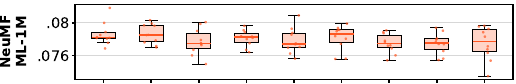}
    \end{subfigure}\\[-1mm]
    \begin{subfigure}{1\linewidth}
    \includegraphics[width=1\linewidth]{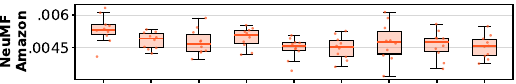}
    \end{subfigure}\\[-1mm]
    \begin{subfigure}{1\linewidth}
    \includegraphics[width=1\linewidth]{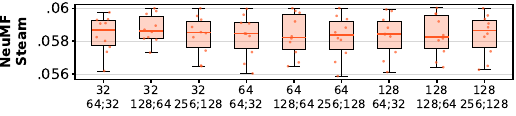}
    \end{subfigure}\\[-1mm]
    \begin{subfigure}{1\linewidth}
    \includegraphics[width=1\linewidth]{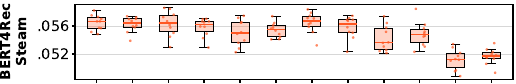}
    \end{subfigure}\\[-1mm]
    \begin{subfigure}{1\linewidth}
    \includegraphics[width=1\linewidth]{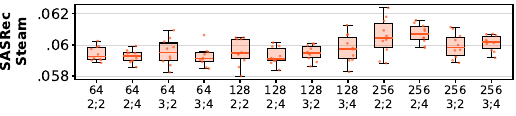}
    \end{subfigure}
    \caption{Per-seed test NDCG@10: NeuMF on the three datasets, then BERT4Rec and SASRec on Steam. Boxplots show mean NDCG@10 across training seeds. Labels on each group's bottom panel: MF embedding size over MLP layer sizes (NeuMF); hidden layers' size; number of heads (BERT4Rec/SASRec).\label{fig:metric_sensitivity}}
    \end{minipage}\hfill
    \begin{minipage}[t]{0.495\textwidth}
    \vspace{0pt}%
    \centering
    \begin{subfigure}{1\linewidth}
    \includegraphics[width=1\linewidth]{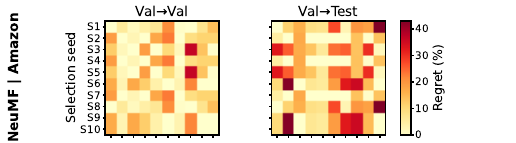}
    \end{subfigure}\\[-0.5mm]
    \begin{subfigure}{1\linewidth}
    \includegraphics[width=1\linewidth]{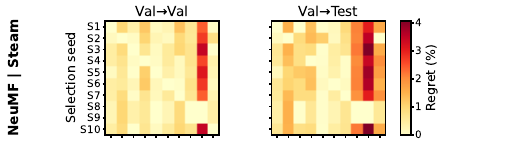}
    \end{subfigure}\\[-0.5mm]
    \begin{subfigure}{1\linewidth}
    \includegraphics[width=1\linewidth]{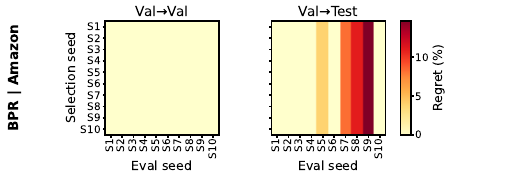}
    \end{subfigure}\\[-0.5mm]
    \caption{Performance gaps from validation-based configuration selection. Rows: seed used to select the best validation configuration; columns: evaluation seed. Colour encodes relative regret (\%), with one shared scale per row.\label{fig:model_stability_cross_regret}}
    \end{minipage}
\end{figure*}

\paragraph{\ul{Metric-level sensitivity}}
Figure~\ref{fig:metric_sensitivity} illustrates metric-level sensitivity for NeuMF, SASRec and BERT4Rec, using test nDCG@10 obtained by each seed for every hyperparameter configuration. The contrast shows that seed variation can take different forms.
Across datasets, training-seed effects are often detectable at the user-score level, but their practical interpretation varies. On ML-1M, Friedman tests show clear metric-level sensitivity for BERT4Rec and SASRec, partial sensitivity for NeuMF, and limited sensitivity for BPR. Steam shows the strongest statistical signal, with Friedman tests rejecting the null for all configurations of all models on validation and test. However, statistical significance does not necessarily imply model-selection instability. For BERT4Rec and SASRec, seed effects often occur within clearly separated configuration regions, while NeuMF is the clearest flatter case, with stronger overlap across configurations. Overall, \textit{metric-level sensitivity is a warning signal, but whether it affects experimental conclusions depends on the size of seed-induced variation relative to the gaps between configurations.}

\vspace{-0.15cm}\paragraph{\ul{Model-selection stability}}
Figure~\ref{fig:model_stability_cross_regret} shows three representative performance-gap cases: NeuMF on Amazon, where both validation--validation and validation--test gaps are high; NeuMF on Steam, where both gaps remain low; and BPR on Amazon, where validation--validation gaps are zero but validation--test gaps are non-zero. Overall, the performance-gap analysis shows that seed variation can affect model selection in different ways. In some cases, different seeds select different validation configurations, but the loss remains small because the competing configurations perform similarly (e.g., NeuMF on Steam). In other cases, different seeds lead to large gaps between the validation-selected configuration and the best observed configuration (e.g., NeuMF and BERT4Rec on Amazon). Finally, validation selection can be stable, but the selected configuration may not transfer as well to test (e.g., BPR on Amazon). Thus, \textit{model-selection stability depends not only on whether the selected configuration changes across seeds, but also on the performance gap and on the alignment between validation and test behavior.}

\vspace{-0.15cm}\paragraph{\ul{Validation--test consistency}}
Results show that validation--test transfer is both dataset- and model-dependent. On ML-1M, validation comparisons are generally informative, although not uniformly so. BERT4Rec and SASRec show mostly high consistency, BPR shows moderate-to-high consistency, and NeuMF is the weakest case. Amazon shows a weaker and more heterogeneous transfer pattern, where several validation winners do not remain better on test. This helps explain the larger validation--test performance gaps observed. Instability comes not only from different seeds selecting different configurations, but also from validation rankings being less reliably preserved on test. BPR provides a useful contrast, as validation selection can be stable while the selected configuration still fails to transfer consistently to test. Steam shows the opposite tendency. Validation comparisons usually transfer well to test, especially for NeuMF, although SASRec remains more heterogeneous. Overall, validation-based model selection depends not only on the stability of the selected configuration across seeds, but also on whether validation comparisons preserve test ordering. Thus, \textit{high seed sensitivity does not necessarily imply large selection losses when competing configurations perform similarly, but weak validation--test transfer can make validation-based choices unreliable even when validation selection appears stable.}

\vspace{-0.15cm}\paragraph{\ul{Recommendation-list agreement}}

\begin{figure*}[t]
    \centering
    \begin{subfigure}{0.360\textwidth}
    \includegraphics[width=1\linewidth]{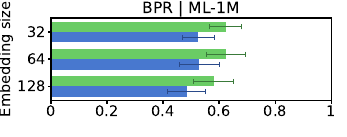}
    \end{subfigure}%
    \begin{subfigure}{0.315\textwidth}
    \includegraphics[width=1\linewidth]{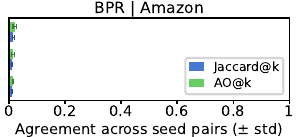}
    \end{subfigure}%
    \begin{subfigure}{0.315\textwidth}
    \includegraphics[width=1\linewidth]{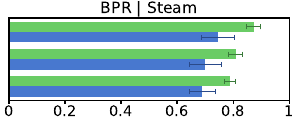}
    \end{subfigure}
    \caption{Cross-seed recommendation-list agreement for BPR on the test split.\label{fig:list_agreement}}
\end{figure*}

\vspace{-0.1cm}Figure~\ref{fig:list_agreement} illustrates this analysis for BPR on the three datasets, using Jaccard@$k$ and AO@$k$ agreement across pairs of seeds on the test split. The contrast is substantial, BPR produces moderately similar lists on ML-1M, almost disjoint lists on Amazon All Beauty, and highly consistent lists on Steam. Across datasets, AO@$k$ is consistently higher than Jaccard@$k$, indicating that different seeds preserve more agreement near the top of the ranking than across the full top-$k$ set. However, agreement remains below $1$ in all cases, so repeated runs are not list-invariant. List-level stability is also strongly dataset-dependent. On ML-1M, agreement is moderate to high, with SASRec producing the most stable lists and NeuMF the least stable ones. On Amazon, stability is weakest, with BPR producing almost no overlap across seeds and BERT4Rec also showing low agreement, while SASRec is substantially more stable. On Steam, agreement is generally high across models, especially for BPR, SASRec, and NeuMF. 
Overall, \textit{these results show that metric-level sensitivity, model-selection instability, and list-level instability are related but distinct effects of training seeds. Importantly, similar aggregate performance may still correspond to different recommendation lists, affecting user exposure.}

\vspace{-0.3cm}\section{Conclusions}

\vspace{-0.15cm}This work shows that training seeds can affect recommender evaluation at several levels. Across datasets and models, changing only the training seed can produce detectable differences in user-level scores, influence validation-based configuration choices, and change the top-$k$ recommendation lists. 
Nonetheless, these effects do not always have the same practical meaning. Some are statistically detectable without substantially changing model-selection outcomes, while others reveal weaker validation--test transfer or different lists under similar aggregate performance.
Overall, these results suggest that training seeds should be treated as part of the evaluation protocol rather than as incidental implementation noise. Reporting seed variation is especially important when configurations are close in performance, when validation and test behavior differ, or when conclusions rely on a single aggregate metric.

This study has several limitations. 
First, the training seed is treated as the seed of the full training run. Thus, the analysis captures run-to-run variability, but does not attribute this variability to a specific stochastic mechanism such as initialization, mini-batch ordering, dropout, latent sampling or training-time negative sampling.
Second, data partition is fixed by design, which allows us to focus on training-seed variation. 
Future work should disentangle individual sources of training randomness, jointly analyze data-split, training and evaluation-sampling seeds, and extend the analysis to additional models, datasets, and beyond accuracy outcomes.

\clearpage

\bibliographystyle{ACM-Reference-Format}
\bibliography{references}

\end{document}